\documentclass[showpacs,amsmath,amssymb]{revtex4}

\begin{document}

\title{Algebraic analysis of quantum search with pure and mixed states}

\author{Daniel Shapira}
\author{Yishai Shimoni}
\author{Ofer Biham}
\affiliation{Racah Institute of Physics, The Hebrew University, 
Jerusalem 91904, Israel}

\newpage

\begin{abstract}

An algebraic analysis of 
Grover's quantum search algorithm
is presented 
for the case in which the initial state is an
arbitrary pure quantum state 
$| \psi \rangle$ of $n$ qubits. 
This approach reveals the geometrical structure of the quantum search process,
which turns out to be confined to a four-dimensional subspace of the
Hilbert space.
It unifies and generalizes earlier results on the 
time evolution of the amplitudes during the search,
the optimal number of iterations
and the success probability.
Furthermore, it enables a direct generalization to the case
in which the initial state is a mixed state, providing
an exact formula for the success probability.
\end{abstract}

\pacs{PACS: 03.67.Lx, 89.70.+c}

\maketitle

\section{Introduction}

Grover's quantum search algorithm 
\cite{Grover1996,Grover1997a} 
exemplifies
the potential speed-up offered by quantum computers. 
It also provides a
laboratory for the analysis of quantum algorithms and their 
implementation. 
The problem addressed by Grover's algorithm can be viewed as trying to 
find a marked element in an unsorted database of size $N$. 
While a classical computer would need, on average, 
$N/2$ database queries (and $N$ queries in the worst case)  
to solve this problem,
a quantum computer 
using Grover's algorithm, would 
accomplish the same task using merely 
$O(\sqrt{N})$ queries. 
This proves the enhanced power of
quantum computers compared to classical ones for a whole class of
oracle-based problems, for which the bound 
on the efficiency of classical
algorithms is known. 
Moreover, it was shown 
\cite{Zalka1999} 
that Grover's algorithm is as 
efficient as theoretically possible 
\cite{Bennett1997}.   
A variety of applications were developed, in which the
algorithm is used in the solution of other
problems
\cite{Grover1997b,Grover1997c,Terhal1998,Brassard1998,Cerf2000,Grover2000,Carlini1999}.

Several generalizations of Grover's original
algorithm have been developed. 
The case in which there are several marked 
states was studied in 
Refs.~\cite{Boyer1996,Boyer1998}. 
It was shown that when there are
$r$ marked states, 
Grover's algorithm can find one of them
after 
$T=O(\sqrt{N/r})$ queries.
A further generalization was obtained 
by allowing 
the replacement of the Hadamard transform, used in the original setting,
by an arbitrary (but constant) unitary
transformation 
\cite{Grover1998,Gingrich2000,Biham2001},
as well as by the replacement of the $\pi$ inversion 
by an arbitrary (but constant) phase rotation 
\cite{Long1999a,Long1999b}. 
Another generalization was obtained 
by allowing the replacement of the 
uniform superposition of all basis states,
used as the initial state of the algorithm
in the original setting, 
by an arbitrary pure 
\cite{Biron1997,Biham1999} 
or mixed
\cite{Biham2002e}
quantum state.
It was shown that the optimal time to perform the
measurement that concludes the operation of the algorithm
is independent of the initial state. 
However, the probability of success, 
$P_{\rm s}$, 
is
reduced, and its value depends on
the initial state.
An explicit expression for $P_s$ 
in terms of the amplitudes of the initial state
was found 
\cite{Biham2003}.
This generalization provides an operational measure of
entanglement of pure multi-partite quantum states
\cite{Miyake2001,Biham2002,Shimoni2004}.

In this paper we introduce an algebraic approach to 
the analysis of 
Grover's quantum search algorithm
with an arbitrary initial quantum state.
This approach reveals the geometrical structure of the search process,
which turns out to be confined to a four dimensional subspace of the
Hilbert space.
This approach 
unifies and generalizes earlier results on the 
time evolution of the amplitudes during the search,
the optimal number of iterations
and the success probability.
Furthermore, it enables a direct generalization to the case
in which the initial state is a mixed state, providing
an exact formula for the success probability.

The paper is organized as follows. 
In Sec. II we briefly describe the algorithm. 
The algebraic analysis is presented in Sec. III
for the general case that involves several marked
states with an arbitrary pure state as the initial state.
Special cases such as the case of a single marked state 
are considered in Sec. IV.
The generalization to mixed initial states is presented
in Sec. V.
The results are summarized in Sec. VI.
The detailed calculation of the success probability of the
algorithm is given in the Appendix.
 
\section{The quantum search algorithm} 
\label{sec:algorithm} 
 
Consider a search space $D$ containing $N$ elements.  We assume, for 
convenience, that $N = 2^n$, where $n$ is an integer. 
The elements of $D$ are represented using an $n$-qubit {\em register} 
containing the indices, $i=0,\dots,N-1$.  We assume that a subset of 
$r$ elements in the search space are marked, that is, they are 
solutions of the search problem.  The distinction between the marked 
and unmarked elements can be expressed by a suitable function, 
$f: D \rightarrow \{0,1\}$, 
such that $f=1$ for the marked elements, and $f=0$ for the rest. 
The search for a marked element now becomes a search for an element 
for which $f=1$.  To solve this 
problem on a classical computer one needs to evaluate $f$ for each 
element, one by one, until a marked state is found.  Thus, on average, 
$N/2$ evaluations of $f$ are required and $N$ in the worst case.
For a quantum computer, on which
$f$ to be evaluated 
\emph{coherently}, 
it was shown that a sequence of unitary operations 
called Grover's algorithm
can locate a marked element using only $O(\sqrt{N/r})$ coherent 
queries of $f$ \cite{Grover1996,Grover1997a}.  
 
To describe the operation of the quantum search algorithm we first 
introduce a register, 
$\left| i \right\rangle = 
\left| i_{1} \ldots   i_{n} \right\rangle$, 
of $n$ qubits, 
and an 
\emph{ancilla} qubit, 
$|q\rangle$, 
to be used in the computation.  
We also introduce a 
\emph{quantum oracle}, 
a unitary operator $\hat{O}$, 
which has the ability to 
\emph{recognize} 
solutions to the search 
problem.  
The oracle performs the following 
unitary operation on computational basis states of the register, 
$\left| i \right\rangle$, 
and the ancilla, 
$\left| q \right\rangle$:
 
\begin{equation}
\hat{O} \left| i \right\rangle \left| q \right\rangle = 
\left| i \right\rangle   \left| q \oplus f(i) \right\rangle 
\label{eq:bborac} 
\end{equation} 

\noindent
where $\oplus$ denotes addition modulo 2. 
The oracle recognizes marked states in the sense that if 
$| i \rangle$ 
is a marked element of the search space, 
namely $f(i) = 1$, 
the oracle flips the ancilla qubit from 
$\left| 0 \right\rangle$ 
to 
$\left| 1 \right\rangle$ 
and vice versa, 
while for unmarked states the ancilla is unchanged.  
The ancilla qubit is initially 
set to the state 
$| - \rangle_q =  (\left| 0 \right> - \left| 1 \right>)/\sqrt{2}$.  
With this choice, the action of the oracle is 
$\hat{O} |i\rangle |-\rangle_q 
= (-1)^{f(i)} |i\rangle |-\rangle_q$.
Thus, the only effect of the oracle is to apply a phase of $-1$ if $x$ 
is a marked basis state, and no phase change if $x$ is unmarked.  
The state of the ancilla does not change. 

Grover's search algorithm may be described as follows: 
Given an oracle $\hat{O}$, whose action 
is defined by Eq.~(\ref{eq:bborac})
and $n+1$ qubits in the 
state $|0\rangle^{\otimes n}|0\rangle_q$,
the following procedure is performed:

\begin{enumerate} 
\item \label{en:initst} Initialization: 
Apply a Hadamard gate    
$\hat{W} = \frac{1}{\sqrt2}
\left(
\begin{smallmatrix} 
1 & 1 \\ 
1 & -1 
\end{smallmatrix}
\right)$ 
to each qubit in the register, 
and the gate $\hat{W}\hat{X}$ to the ancilla, 
where 
$\hat{X}=
\left(
\begin{smallmatrix} 
0 & 1 \\ 
1 & 0 
\end{smallmatrix}
\right)$ 
is the {\sc not} gate,  
and we write matrices with respect to the computational basis 
($|0\rangle,|1\rangle$).  
The resulting state is 
$|\eta \rangle |-\rangle_q$, 
where

\begin{equation}
| \eta \rangle = \frac{1}{\sqrt{N}} \sum_{i=0}^{N-1} |i\rangle. 
\label{eq:eta} 
\end{equation} 

\noindent
\label{init:item} 
\item Grover Iterations: Repeat the following operation $\tau$ times 
  (where $\tau$ is given below). 
\begin{enumerate} 
\item \label{en:rot1} Apply the oracle, which has the effect of 
rotating the marked states by a phase of $\pi$ radians.  Since the 
ancilla is always in the state $|-\rangle_q$ the effect 
of this operation 
may be described by the unitary operator 
$\hat{I}_M = \hat{I} - 2 \sum_{m \in M} | m \rangle \langle m |$, 
acting only on the register,
where $\hat{I}$ is the identity operator. 

\item (i) apply the 
Hadamard gate on each qubit in the register; 
(ii) apply the operator 
$\hat{I}_{0} = \hat{I} - 2 |0\rangle \langle 0|$
which   
rotates the 
$\left| 00 \ldots 0 \right\rangle$ state  
of the register by a phase of $\pi$ 
radians.  
(iii) Apply the Hadamard gate again on each qubit in the register. 

The resulting operation is
$- \hat{W} \hat{I}_{0} \hat{W} = 
    -\hat{I} + 2 | \eta \rangle \langle \eta |$.
When this operator is applied on the state 
$\sum_i a_i | i \rangle$
it results in the state 
$\sum_i (2 \bar{a} - a_i) | i \rangle$,
where $\bar a = \sum_i a_i/N$. 
Thus, each amplitude is rotated by $\pi$ around the
average of all amplitides of the quantum state.
\end{enumerate} 

\item  Measure the register in the computational basis. 
\end{enumerate}

The combined operation on the register
in one Grover iteration 
is given by 
$\hat{Q} = - \hat{W} \hat{I}_{0} \hat{W} \hat{I}_M$. 
The optimal number of iterations before the measurement is

\begin{equation}
\tau = \left\lfloor
\frac{\pi}{4} \sqrt{\frac{N}{r}}
\right\rfloor, 
\label{eq:optit} 
\end{equation} 

\noindent
where $\lfloor x \rfloor$
is the largest integer which is smaller than $x$
\cite{Grover1997a,Boyer1998,Zalka1999}. 
Moreover, at this optimal time a marked state
can be found with almost certainty, or more 
precisely with probability
$P_{\rm s} = 1 - O \left( 1/\sqrt{N} \right)$. 
With this performance,
Grover's algorithm was found to be 
optimal in the sense that it is as efficient as 
theoretically possible
\cite{Bennett1997}. 
Note that the probability 
$P_s \approx 1$ 
can be achieved only 
for specific initial states such as the one 
produced in step~$1$ of the algorithm above.  
If this initial state is replaced by 
an arbitrary quantum state, the
probability of succes, $P_{\rm s}$, is reduced
\cite{Biron1997,Biham1999}. 

The time evolution of the amplitudes of the marked
and unmarked states during Grover's iterations
was studied in Ref.
\cite{Biham1999}
for an arbitrary pure initial state
$| \psi \rangle$.
In particular, the optimal number of iterations
and the success probability
were calculated, and found to depend on 
the specific choice of the set of marked states.
Obviously, in a search process, the marked states are
not known.
Thus, the success probability should be averaged over all
possible choices of the set of $r$ marked states
\cite{Biham2003}.
The results are that for any initial state
$| \psi \rangle$
the optimal number of iterations is given by
Eq.
(\ref{eq:optit}).
The success probability,
$P_s$ can be expressed in terms of the
average of all amplitudes of the state
$| \psi \rangle$.
In the next Section we introduce the algebraic approach to
the analysis of Grover's search algorithm.

\section{Algebraic analysis of the quantum search process}

Consider a search using Grover's algorithm, 
for one of $r$ marked states 
in a space of $N=2^n$ 
computational basis states,
where $n$ is the number of qubits in the register. 
The initial state is 

\begin{equation}
|\psi \rangle = \sum_{i=0}^{N-1} a_i |i \rangle 
\end{equation}

\noindent
where
$a_i$ is the amplitude of the basis state
$|i \rangle$. 
Denote the set of indices of the marked states by 
${\cal M}$.
The amplitudes of the marked states will thus be
$a_m$,  $m \in {\cal M}$.
The complementary set, of unmarked states is denoted by 
${\cal U}$, namely the amplitudes of the unmarked states are 
$a_u$, $u \in {\cal U}$.
Thus, for a given choice of the set of marked states, 
the initial state $|\psi \rangle$ 
can be expressed by

\begin{equation}
|\psi \rangle   =  
\sum_{m \in {\cal M}} a_m |m \rangle + 
\sum_{u \in {\cal U}} a_u |u \rangle.
\label{psi_a}
\end{equation}

\noindent
The amplitudes 
$a_m$
and
$a_u$
satisfy:

\begin{equation}
 \sum_{m \in {\cal M}} {|a_m|}^2 + 
\sum_{u \in {\cal U}} {|a_u|}^2 = 1.
\label{a_normalization}
\end{equation}

\noindent
Their averages are given by
 
\begin{equation}
{\bar{a}}_M = \frac{1}{r}  \sum_{m \in {\cal M}} a_m 
\label{aM}
\end{equation}

\noindent
for the marked states, and 

\begin{equation}
{\bar{a}}_U = \frac{1}{N-r}  \sum_{u \in {\cal U}} a_u 
\label{aU}
\end{equation}

\noindent
for the unmarked states.    

\subsection{Construction of a four-dimensional subspace}

The initial state
$| \psi \rangle$
can be projected onto the
subspaces 
spanned by
the marked and unmarked basis states,
giving rise to the normalized projections 

\begin{equation}
|\phi_M \rangle  = \frac{1}{\sqrt{P_0}} \sum_{m \in {\cal M}} a_m |m \rangle 
\label{phi_M}
\end{equation}

\noindent
and 

\begin{equation}
|\phi_U \rangle = \frac{1} {\sqrt{1-P_0}} \sum_{u \in {\cal U}}a_u |u \rangle, 
\label{phi_U}
\end{equation}

\noindent
respectively,
where
\begin{equation}
P_0 = \sum_{m \in {\cal M}} {|a_m|}^2.
\label{P_0}
\end{equation}

\noindent
The state 
$|\psi \rangle$ 
can now be written in the form

\begin{equation}
|\psi \rangle = \sqrt{P_0} |\phi_M \rangle + 
\sqrt{1-P_0} |\phi_U \rangle, 
\label{psi_phi}
\end{equation}

\noindent
where
$|\phi_M \rangle$ 
and 
$|\phi_U \rangle$ 
are orthogonal to each other. 
Similarly, the equal superposition state
$| \eta \rangle$,
can be expressed in the form

\begin{equation}
|\eta \rangle = \sqrt{\frac{r}{N}} |\eta_M \rangle + 
\sqrt{1 - \frac{r}{N}} |\eta_U \rangle, 
\label{eta_theta}
\end{equation}

\noindent
where

\begin{equation}
|\eta_M \rangle = \frac{1}{\sqrt{r}}  
 \sum_{m \in {\cal M}} |m \rangle 
\label{theta_M}
\end{equation}

\noindent
and 

\begin{equation}
|\eta_U \rangle = \frac{1}{\sqrt{N-r}}  
\sum_{u \in {\cal U}} |u \rangle. 
\label{theta_U}
\end{equation}

\noindent
From now on we will refer to the plane spanned by
$|\eta_M \rangle$ 
and
$|\eta_U \rangle$
as the {\it Grover plane}. 
Using the Gram-Schmidt procedure we extract from
$|\phi_M \rangle$ 
and 
$|\phi_U \rangle$ 
two new states
$|\psi_M \rangle$ 
and 
$|\psi_U \rangle$, 
that are perpendicular to
$|\eta_M \rangle$ 
and 
$|\eta_U \rangle$. 
These states are 
given by

\begin{equation}
|\psi_M \rangle = \frac{|\phi_M \rangle - 
\langle \eta_M |\phi_M \rangle |\eta_M \rangle}
{\sqrt{1 - {|\langle \eta_M |\phi_M \rangle|}^2}} 
\label{psi_M_Gram_Scmidt}
\end{equation}

\noindent
and 

\begin{equation}
|\psi_U \rangle = \frac{|\phi_U \rangle - 
\langle \eta_U |\phi_U \rangle |\eta_U \rangle}
{\sqrt{1 - {|\langle \eta_U |\phi_U \rangle|}^2}}. 
\label{psi_U_Gram_Scmidt}
\end{equation}

\noindent
Using the fact that

\begin{eqnarray}
\langle \eta_M |\phi_M \rangle &=& 
\sqrt{\frac{r}{P_0}} {\bar{a}}_M \nonumber \\
\langle \eta_U |\phi_U \rangle &=& 
\sqrt{\frac{N-r}{1-P_0}} {\bar{a}}_U,
\end{eqnarray}

\noindent
we can write 
$|\psi_M \rangle$
and 
$|\psi_U \rangle$
more explicitly as

\begin{eqnarray}
|\psi_M \rangle &=& 
\frac{ \left( \sqrt{P_0} |\phi_M \rangle - 
\sqrt{r}{\bar{a}}_M |\eta_M \rangle \right) }
{\sqrt{P_0 - r{|{\bar{a}}_M|}^2}} 
 \nonumber \\
|\psi_U \rangle &=& 
\frac{ \left( \sqrt{1-P_0} |\phi_U \rangle - 
\sqrt{N-r}{\bar{a}}_U |\eta_U \rangle \right) }
{\sqrt{1 - P_0 - (N-r){|{\bar{a}}_U|}^2}}. 
\label{psi_U}
\end{eqnarray}

\noindent
Thus, Eq. (\ref{psi_phi}) now takes the form

\begin{equation}
|\psi \rangle  =  \sqrt{P_0 - r{|{\bar{a}}_M|}^2} |\psi_M \rangle 
+ \sqrt{1 - P_0 - (N-r){|{\bar{a}}_U|}^2} |\psi_U \rangle  
+ \sqrt{N-r}{\bar{a}}_U |\eta_U \rangle
+ \sqrt{r}{\bar{a}}_M |\eta_M \rangle,
\label{psi_4}
\end{equation}

\noindent
where
the state vectors 
$|\psi_M \rangle$, 
$|\psi_U \rangle$, 
$|\eta_U \rangle$
and 
$|\eta_M \rangle$ 
define a set of four orthonormal basis vectors. 
In particular,
$|\psi_M\rangle$ 
and 
$|\eta_M\rangle$
span the subspace of marked states,
while 
$|\psi_U\rangle$ 
and 
$|\eta_U\rangle$ 
span the subspace of unmarked states. 
Also,
$|\psi_M\rangle$ 
and 
$|\psi_U\rangle$ 
are perpendicular to 
$|\eta\rangle$. 
We will show below that for any initial state 
$|\psi \rangle$, 
iterations of the quantum search always preserve the subspace 
spanned by these four vectors. 

\subsection{The time evolution of the state vector}

Denoting the Grover iteration by
$\hat{Q}$, 
the state of the register after $t$ iterations is 

\begin{equation}
|g(t) \rangle = {\hat{Q}}^t |\psi \rangle.
\label{|gt>}
\end{equation}

\noindent
Using the structure of the operator $\hat{Q}$
presented above, we obtain that for any state 
$| \psi \rangle$

\begin{equation}
\hat{Q} |\psi\rangle =
 -|\psi\rangle + 2 \left( \langle\eta|\psi\rangle -
2\sum_{m \in{\cal M}} \langle\eta|m\rangle\langle m|\psi\rangle \right)
|\eta\rangle 
+ 2\sum_{m \in{\cal M}} \langle m|\psi\rangle|m\rangle.
\label{Qpsi}
\end{equation}

\noindent
Applying $\hat Q$
on the vectors that span the four-dimensional subspace we obtain

\begin{eqnarray}
\hat{Q} |\psi_M\rangle 
&=& |\psi_M\rangle
\nonumber \\
\hat{Q} |\psi_U\rangle 
&=& -|\psi_U\rangle
\nonumber \\
\hat{Q} |\eta_U\rangle 
&=& \left(1-\frac{2r}{N}\right)|\eta_U \rangle + 
2\sqrt{\frac{r}{N}\left(1-\frac{r}{N}\right)} |\eta_M \rangle
\nonumber \\
\hat{Q} |\eta_M\rangle 
&=& -2\sqrt{\frac{r}{N}\left(1-\frac{r}{N}\right)} |\eta_U \rangle +
\left(1-\frac{2r}{N}\right)|\eta_M \rangle. 
\label{Qtheta_M}
\end{eqnarray}

\noindent
The Grover iteration $\hat{Q}$ acts as a linear 
transformation within a four dimensional vector space, 
spanned by  $|\psi_M\rangle$, $|\psi_U\rangle$, $|\eta_U\rangle$ 
and $|\eta_M\rangle$. 
For the analysis presented below it is convenient to 
use the vector representation

\begin{equation}
|\psi_M\rangle \equiv 
\left( \begin{array}{c} 1 \\ 0 \\ 0 \\ 0 \end{array} \right) 
\hspace{.4in} 
|\psi_U\rangle \equiv 
\left( \begin{array}{c} 0 \\ 1 \\ 0 \\ 0 \end{array} \right) 
\hspace{.4in} 
|\eta_U\rangle \equiv 
\left( \begin{array}{c} 0 \\ 0 \\ 1 \\ 0 \end{array} \right) 
\hspace{.4in} 
|\eta_M\rangle \equiv 
\left( \begin{array}{c} 0 \\ 0 \\ 0 \\ 1 \end{array} \right), 
\label{4d_vectors}
\end{equation}

\noindent
in which, according to 
Eq. (\ref{psi_4})

\begin{equation}
|\psi \rangle \equiv 
\left( \begin{array}{c} 
\sqrt{P_0 - r{|{\bar{a}}_M|}^2} \\
\sqrt{1 - P_0 - (N-r){|{\bar{a}}_U|}^2} \\
\sqrt{N-r}{\bar{a}}_U \\
\sqrt{r}{\bar{a}}_M 
\end{array} 
\right).
\label{psi_4_vector}
\end{equation}

\noindent
The matrix representation of $\hat{Q}$ is:

\begin{equation}
\hat{Q} = \left( \begin{array}{cccc} 
1 & 0 & 0 & 0 \\
0 & -1 & 0 & 0 \\
0 & 0 & \cos \omega & -\sin \omega \\
0 & 0 & \sin \omega & \cos \omega  
\end{array} \right)
\label{Q_matrix}
\end{equation}

\noindent
where 

\begin{equation}
\cos \omega = 1 - \frac{2r}{N} 
\label{cosw}
\end{equation}

\noindent
and 

\begin{equation}
\sin \omega = 2\sqrt{\frac{r}{N}\left(1-\frac{r}{N}\right)}. 
\label{sinw}
\end{equation}   

\noindent
Therefore,

\begin{equation}
{\hat{Q}}^t = \left( \begin{array}{cccc} 
1 & 0 & 0 & 0 \\
0 & {(-1)}^t & 0 & 0 \\
0 & 0 & \cos (\omega t) & -\sin (\omega t) \\
0 & 0 & \sin (\omega t) & \cos (\omega t)  
\end{array} \right)
\label{Qt_matrix}
\end{equation}

\noindent
and the state of the register after $t$ iterations is

\begin{eqnarray}
|g(t) \rangle & = & 
\sqrt{P_0 - r{|{\bar{a}}_M|}^2} \ |\psi_M \rangle + 
{(-1)}^t 
\sqrt{1 - P_0 - (N-r){|{\bar{a}}_U|}^2} \  |\psi_U \rangle 
\nonumber \\
& & + \left( \sqrt{N-r}{\bar{a}}_U\cos(\omega t)
-\sqrt{r}{\bar{a}}_M\sin(\omega t)\right) 
|\eta_U \rangle  
\nonumber \\
& & + \left( \sqrt{N-r}{\bar{a}}_U \sin(\omega t) 
+ \sqrt{r}{\bar{a}}_M \cos(\omega t) \right) 
|\eta_M \rangle. 
\label{eq:gt4}
\end{eqnarray}

\noindent
The four dimensional space in which the dynamics is confined
includes the {\it plane of marked states},
spanned by 
$|\psi_M \rangle$ 
and 
$|\eta_M \rangle$ 
as well as the
{\it plane of unmarked states}
spanned by 
$|\psi_U \rangle$ 
and 
$|\eta_U \rangle$. 

\subsection{The success probability}

The probability of success $P_{s}(t)$
of a measurement taken  
after $t$ iterations is given by 
the projection of $|g(t)\rangle$ 
on the plane of marked states:   

\begin{eqnarray}
P_s(t) & = & 
P_0 + \frac{1}{2} \left[(N-r){|{\bar{a}}_U|}^2 
- r{|{\bar{a}}_M|}^2\right] 
\nonumber \\
& & - \frac{1}{2} \left[(N-r){|{\bar{a}}_U|}^2
-r{|{\bar{a}}_M|}^2\right] \cos(2\omega t)
\nonumber \\
& & + \frac{1}{2} \sqrt{r(N-r)} 
({{\bar{a}}_U}^*{\bar{a}}_M + {{\bar{a}}_M}^*{\bar{a}}_U ) \sin(2\omega t). 
\label{Pst}
\end{eqnarray}

\noindent From Eq.
(\ref{eq:gt4}) 
it is easy to see that
$| g(t) \rangle$
exhibits a rotation
with angular velocity 
$\omega$  
in the
{\it Grover plane}
spanned by 
$|\eta_U \rangle$ 
and 
$|\eta_M \rangle$.
This rotation causes an exchange of probability between
the planes of marked and unmarked states.
Thus, the success probability is limited by the projection 
of the initial state 
$|\psi \rangle$ 
on the Grover plane. 
The projection of  
$|\psi \rangle$ 
along
$|\psi_U \rangle$, 
which is perpendicular to the Grover plane,
represents a lost probability that cannot be
transformed into the plane of the marked states. 
This provides an upper bound 
on the success probability  
$P_{s}(t)$,
of the form

\begin{equation}
P_{\rm max} = P_0 + (N-r){|{\bar{a}}_U|}^2. 
\label{maxPst}
\end{equation}

\noindent
Similarly, 
a lower bound on the success probability

\begin{equation}
P_{\rm min} = P_0 - r{|{\bar{a}}_M|}^2 
\label{minPst}
\end{equation}

\noindent
can be obtained using the projection in the direction of
$|\psi_M \rangle$.
Furthermore, by using the explicit forms of the 
states that span the four dimensional subspace 
and Eq.~(\ref{eq:gt4}), 
the following expression for 
the temporal evolution of the 
state vector is obtained

\begin{equation}
|g(t)\rangle = \sum_{m \in {\cal M}} 
\left( \bar{k}(t) + \Delta a_m \right) |m\rangle 
+ \sum_{u \in {\cal U}} \left(\bar{l}(t) 
+ {(-1)}^t \Delta a_u \right) |u\rangle.    
\label{gt_k_l}
\end{equation}

\noindent
The variables

\begin{equation}
\bar{k}(t) = \sqrt{\frac{N-r}{r}} \ \bar{a}_U \sin(\omega t) 
+ \bar{a}_M \cos(\omega t) 
\label{k} 
\end{equation}

\noindent
and 

\begin{equation}
\bar{l}(t) = \bar{a}_U \cos(\omega t) -
\sqrt{\frac{r}{N-r}} \bar{a}_M \sin(\omega t) 
\label{l} 
\end{equation}

\noindent
are the average amplitudes, 
after $t$ iterations,
of the marked and unmarked states,
respectively.
Also,

\begin{eqnarray}
\Delta a_m &=& a_m - {\bar{a}}_M, \ \ \ \      {m \in {\cal M}} 
\nonumber \\
\Delta a_u &=& a_u - {\bar{a}}_U, \ \ \ \ \      {u \in {\cal U}}  
\end{eqnarray}

\noindent
represent the initial deviations of the
amplitudes $a_m$ and $a_u$ from the averages of 
the marked and unmarked amplitudes,
respectively
\cite{Biham1999}.

The analysis presented above applies to any specific choice of the
set of marked state. In practice, the set of marked states
is not known. 
Thus, the success probability 
of an actual search
is the average of
$P_s(t)$
over all possible choices of the set of $r$ marked states. 
In the limit of a large search space, where
$r \ll N$, it
takes the form
(see Appendix):

\begin{equation}
\langle P_s(t) \rangle =   N{|\bar{a}|}^2 {\sin}^2 [\omega (t+1/2)] 
+ O\left({r/N}\right)
\label{<pst>_r_ll_N},
\end{equation}

\noindent
where

\begin{equation}
\bar{a} = \frac{1}{N} \sum_{i=0}^{N-1} a_i 
\label{bar_a}
\end{equation}

\noindent
is the average  
over all the amplitudes of
the initial state 
$|\psi \rangle$. 
The optimal 
number of iterations, $\tau$, is given by Eq.
(\ref{eq:optit}),
namely,
is identical to that obtained for 
the original algorithm.
The probability of success 
of a measurement taken after $\tau$ iterations is 

\begin{equation}
P_{\rm max} = N{|\bar{a}|}^2 + O\left({r/N}\right). 
\label{P}
\end{equation}

\noindent
It can also be expressed by
$P_{\rm max} = |\langle \eta | \psi \rangle|^2  + O({r/N})$
\cite{Biham2003}.

\section{Analysis of special cases}

Below we consider special cases in which the analysis
can be simplified and the dimension of the subspace in which
the Grover iterations operate is reduced.

\subsection{Quantum search with a single marked state}

Consider the quantum search algorithm with an initial state 
$|\psi \rangle$ 
and a single marked state $|m\rangle$. 
In this case
the subspace of the marked states is 
one dimensional. 
Therefore, the normalized projection of the equal
superposition state 
$|\eta\rangle$ 
on the subspace of marked states is given by

\begin{equation}
|\eta_M \rangle = |m \rangle.
\label{thetaM_1}
\end{equation}

\noindent
Since the 
one dimensional subspace of the marked states
does not include any 
perpendicular direction 
$|\psi_M \rangle = 0$.
The average amplitude of the marked states is 
$\bar{a}_M = a_m$, 
where $a_m$ is the amplitude of the marked state 
$|m \rangle$ 
in the initial state 
$|\psi \rangle$. 
Thus, the state
$|g(t) \rangle$ 
of the register after $t$ Grover iterations 
[Eq. (\ref{eq:gt4})] 
takes the form

\begin{eqnarray}
|g(t) \rangle & = & 
\left(\sqrt{N-1}\bar{a}_U\cos(\omega t) - a_m\sin(\omega t) \right)  
|\eta_U \rangle 
+ \left(\sqrt{N-1}\bar{a}_U\sin(\omega t) + a_m\cos(\omega t) \right)  
| m \rangle 
\nonumber \\
& & + {(-1)}^t \sqrt{1-(N-1){|\bar{a}_U|}^2 - {|a_m|}^2} \ |\psi_U \rangle.
\label{gt3}
\end{eqnarray}

\noindent
Clearly,
$|g(t) \rangle $
is confined to the
three dimensional subspace spanned 
by  
$|\eta_U \rangle$, 
$|m \rangle$ 
and 
$|\psi_U \rangle$. 

In case that all the amplitudes $a_i$ are real, 
the dynamics of the quantum search can be viewed
as a rotation of the  
state vector 
$|g(t) \rangle$ 
around the bases of a cylinder,
jumping from one base to the other at
each time step. 
The axis of the cylinder  
is in the direction of
$|\psi_U \rangle$. 
Its bases 
are in the Grover plane, namely
spanned by 
$|\eta_U \rangle$ 
and 
$|m \rangle$. 
The radius of the cylinder is given by 
the projection of 
$|g(t) \rangle$ 
on the Grover plane:

\begin{equation}
R = \sqrt{(N-1){|\bar{a}_U|}^2+{|a_m|}^2}.
\end{equation}

\noindent
The length of the cylinder is twice the size of the projection of 
$|g(t) \rangle$ 
in the direction of 
$|\psi_U \rangle$,
namely 

\begin{equation}
L = 2 \sqrt{1-R^2}.
\end{equation}

\noindent
The dynamics of the quantum 
search consists of rotations of the state vector 
$|g(t) \rangle$ 
around the 
$|\psi_U \rangle$ 
axis, at an angular velocity 
$\omega$, 
combined with switching positions between the
two bases.
At even time steps 
$|g(t) \rangle$ 
points towards 
the upper base, while at odd times
it points towards the lower base. 
The optimal measurement time 
is when the vector 
$|g(t) \rangle$ 
is exactly above (or below) 
the 
$|m \rangle$ axis. 

\subsection{Quantum search with an initial state in the Grover plane}

Consider a quantum search with a certain set of $r$ marked states,
such that the
initial state 
$| \psi \rangle$
is in the Grover plane. 
In this case 
$| \psi \rangle$
can be represented 
by

\begin{equation}
|\psi \rangle = \alpha |\eta_U \rangle + \beta  |\eta_M \rangle
\label{psi_in_Grover_plane}
\end{equation}

\noindent
where 
$\alpha$ 
and 
$\beta$ are complex amplitudes that satisfy
${|\alpha|}^2 + {|\beta|}^2 = 1$.
Note that 
$|\psi \rangle$ 
has no component perpendicular to the Grover 
plane either in the subspace of marked states or in the subspace 
of unmarked states. 
Therefore, 
$|\psi_M \rangle = |\psi_U \rangle = 0$.
The average amplitudes 
of the marked and unmarked basis states
in 
$|\psi \rangle$
are
$\bar{a}_M = \beta / \sqrt{r}$ 
and
$\bar{a}_U = \alpha / \sqrt{N-r}$, 
respectively.
The state vector after $t$ iterations takes the form

\begin{equation}
|g(t) \rangle = 
\left[ \alpha \cos(\omega t) - \beta \sin(\omega t) \right]  |\eta_U \rangle + 
\left[ \alpha \sin(\omega t) + \beta \cos(\omega t) \right]  |\eta_M \rangle,
\label{gt_thetaU_thethaM}
\end{equation}

\noindent
namely, it is confined to the Grover plane, where it rotates 
with angular velocity $\omega$. 
The success probability 
$P_s(t)$ 
oscillates periodically 
between 
$P_{\rm min}=0$
and
$P_{\rm max}=1$,
namely the algorithm is optimal.
For example, 
the equal superposition state 
$|\eta \rangle$ 
is a special case within this category,
where
$\alpha = \sqrt{N-r/N}$ 
and 
$\beta = \sqrt{{r}/{N}}$. 

\subsection{Quantum search with an initial state 
perpendicular to the Grover plane}

Consider an initial state 
$|\psi \rangle$ 
which is perpendicular to the Grover plane,
namely a state that satisfies 
$\langle \eta_M | \psi \rangle = \langle \eta_U | \psi \rangle = 0$.
In this case 
$\bar{a}_M = \bar{a}_U = 0$ and

\begin{equation}
|g(t) \rangle = \sqrt{P_0} |\psi_M \rangle + 
{(-1)}^t \sqrt{1-P_0} |\psi_U \rangle. 
\label{gt_psiM_psiU}
\end{equation}

\noindent
This state vector is confined to 
a cycle of period 2, namely 
$|g(t+2) \rangle = |g(t) \rangle$. 
Recall that Grover's algorithm 
generates a flow of probability from the subspace of unmarked states 
toward the subspace of marked states only within the Grover plane. 
In this case 
$|\psi \rangle$ 
is perpendicular to the Grover 
plane.
Therefore, 
the probability of success remains unchanged,
namely
$P_s(t) = P_0$,
making the quantum search process useless.
Two special cases can be identified. In case
that
$\bar a_U = 0$
and
$a_m = 0$
for all 
$m \in M$,
the success probability 
$P_s(t) = 0$
at any time $t$.
In the opposite case
where
$\bar{a}_M = 0$ 
and 
$a_u = 0$ for all $u \in {\cal U}$,
$P_s(t) = 1$ 
at all times.
 
\section{Quantum search using mixed states}

The quantum search with an initial state which is 
a mixed state was studied before for a specific 
choice of the set of marked states
\cite{Biham2002e}.
Here we extend the analysis to the 
general 
case in which the set of marked states is unknown
and calculate the success probability.
We first analyze the case of a genral mixed states
and then discuss some special cases. 

\subsection{General analysis for arbitrary mixed states}

Consider a quantum search using the 
mixed state 

\begin{equation}
{\hat{\rho}}_0 = \sum_\mu p_\mu |\psi_\mu \rangle \langle \psi_\mu | 
\label{rho0}
\end{equation}

\noindent
of $n$ qubits as the initial state of the register.
In this representation
$p_\mu$ 
is the probability that corresponds to the pure state 
$|\psi_\mu \rangle$ 
in the ensemble
and
$\sum_{\mu} p_{\mu} = 1$.
The pure states take the form

\begin{equation}
|\psi_\mu \rangle = \sum_{i=0}^{N-1} a_{\mu i} |i \rangle. 
\label{psi_mu}
\end{equation}

\noindent
The average amplitude in the state
$|\psi_\mu \rangle$ 
is

\begin{equation}
\bar{a}_\mu = \frac{1}{N} \sum_{i=0}^{N-1} a_{\mu i}. 
\label{a_mu}
\end{equation}

\noindent
The density 
operator after $t$ iterations 
is given by

\begin{equation}
\hat{\rho}(t) = {\hat{Q}}^t {\hat{\rho}}_0 {\hat{Q}}^{t \dagger} 
= \sum_\mu p_\mu |g_\mu(t) \rangle \langle g_\mu(t) |
\label{rhot}
\end{equation}

\noindent
where 
$|g_\mu(t)\rangle = {\hat{Q}}^t |\psi_\mu \rangle$. 
The measurement of the register in the computational basis is represented 
by the operator:

\begin{equation}
\hat{M} = \sum_{i=0}^{N-1} i {\hat{M}}_i, 
\label{hat_M}
\end{equation}

\noindent
where 
${\hat{M}}_i = | i \rangle \langle i |$ 
is the measurement operator 
that corresponds to the outcome $i$. 
The success probability 
of a measurement taken after $t$ iterations
is

\begin{equation} 
P_s(t) = 
\sum_{m \in {\cal M}} {\rm Tr} 
\left({{\hat{M}}_m}^{\dagger} {\hat{M}}_m \hat{\rho}(t)\right) 
= \sum_\mu p_\mu P_s(\psi_\mu,t) 
\label{pst_mix}
\end{equation}

\noindent
where 

\begin{equation}
P_s(\psi_\mu,t) = 
\sum_{m \in {\cal M}} {|\langle m|g_\mu(t)\rangle|}^2. 
\label{Pst_mu}
\end{equation}

\noindent
Averaging the probability of success 
over all possible choices of the set of $r$ marked states,
where $r \ll N$,
gives 

\begin{equation}
\langle P_s(t) \rangle = 
N{|\bar{a}|}^2 {\sin}^2 \left[\omega(t+1/2)\right] + O({r/N}), 
\label{<pst>_mix}
\end{equation} 

\noindent
where

\begin{equation}
{|\bar{a}|}^2 = \sum_\mu p_\mu {|\bar{a}_\mu|}^2. 
\label{a_mix}
\end{equation}

\noindent
Thus, the maximal success probability

\begin{equation}
P_{\rm max} = N |\bar{a}|^2 + O({r/N})
\end{equation}

\noindent
is simply the weighted average of the success
probabilities obtained using the pure states
$| \psi \rangle_{\mu}$
as initial states.
Using 
Eq. 
(\ref{a_mix})
and the fact that 
$|\langle \eta  | \psi_{\mu}  \rangle|^2 =  N {|\bar{a}_\mu|}^2$
we find that
$P_{\rm max} = \langle \eta | {\hat{\rho}}_0 | \eta \rangle$.
This is, in fact, the square of the fidelity of 
${\hat{\rho}}_0$
and 
$| \eta \rangle$,
namely,

\begin{equation}
P_{\rm max} = F^2(| \eta \rangle,{\hat{\rho}}_0).
\end{equation}

\subsection{Search using only part of the qubits in the register}

Consider a quantum register of $n+k$ qubits, which is in
a pure state $| \psi \rangle$.
Suppose that the qubits are divided between two parties
such that Alice gets the first $n$ qubits and Bob
gets the rest of them.
The state of the register can be expressed by

\begin{equation}
|\psi \rangle = \sum_{\mu=0}^{K-1} \sum_{i=0}^{N-1} 
b_{\mu i} |i \rangle_A |\mu \rangle_B 
\label{psi_AB}
\end{equation} 

\noindent
where the subsystem of Alice is spanned by the basis
$|i \rangle_A$, $i=0,1,\dots,N-1$,
where
$N=2^n$,  
while the subsystem of Bob 
is spanned by
$|\mu \rangle_B$, $\mu=0,1,\dots,K-1$, 
where 
$K=2^k$.
The normalization condition is

\begin{equation}
\sum_{\mu=0}^{K-1} \sum_{i=0}^{N-1} {|b_{\mu i}|}^2 = 1
\label{b_normalized}
\end{equation}

\noindent
and 

\begin{equation}
\bar{b} = \frac{1}{NK}\sum_{\mu=0}^{K-1} \sum_{i=0}^{N-1} b_{\mu i} 
\label{b_bar}
\end{equation}

\noindent
is the average amplitude. 
Eq. (\ref{psi_AB}) can be written 
in the form

\begin{equation}
|\psi \rangle =  \sum_{\mu=0}^{K-1} 
c_\mu |\psi_\mu\rangle_A |\mu\rangle_B 
\label{psi_AB_ps_mu}
\end{equation}

\noindent
where 

\begin{equation}
\sum_{\mu=0}^{K-1} {|c_\mu|}^2 = 1. 
\label{c_normalized}
\end{equation} 

\noindent
The pure state 
$|\psi_\mu\rangle_A$ 
of party $A$, that corresponds 
to the computational basis state 
$|\mu\rangle_B$ 
of party $B$ is 
given by 

\begin{equation}
|\psi_\mu\rangle_A =  \sum_{j=0}^{N-1} a_{\mu j} |j \rangle_A  
\label{psi_mu_A}
\end{equation}

\noindent
where
$b_{\mu i} = c_{\mu} a_{\mu i}$.
This state
satisfies the normalization condition

\begin{equation}
\sum_{i=0}^{N-1} {|a_{\mu i}|}^2 = 1 
\label{a_normalized}
\end{equation}

\noindent
and its average amplitude is

\begin{equation}
{\bar{a}}_\mu = \frac{1}{N} \sum_{i=0}^{N-1} a_{\mu i}.   
\label{a_mu_bar}
\end{equation}

\noindent
The measurement statistics for Alice
is given by the reduced density operator:

\begin{equation}
{\hat{\rho}}^A = {\rm Tr}_B {|\psi\rangle\langle\psi|} = 
\sum_{\mu=0}^{K-1}  p_\mu {|\psi_\mu \rangle_{A \  A}\langle \psi_\mu |}
\label{rho_A}
\end{equation}

\noindent
where Tr$_B$ is a trace over the state of Bob and
$p_\mu = {|c_\mu|}^2$.

Consider a quantum search in a search space of size 
$NK$
using the entire register of $n+k$ qubits
in the state 
$| \psi \rangle$.
The success probability is

\begin{equation}
P_{\rm max}^{\rm AB} = NK |\bar b|^2 = \frac{N}{K} 
\left|\sum_{\mu=0}^{K-1} c_{\mu} \bar a_{\mu} \right|^2
+ O\left(\frac{r}{NK}\right). 
\end{equation}

\noindent
Now, consider the case in which Alice performs a quantum search in a 
space of size $N$ using her $n$ qubits in the state
$\hat \rho^A$. The success probability of Alice's
search is

\begin{equation}
P_{\rm max}^{\rm A} = N \sum_{\mu=0}^{K-1} p_{\mu} |\bar a_{\mu}|^2
     = N \sum_{\mu=0}^{K-1}  | c_{\mu} \bar a_{\mu}|^2
+ O({r/N}). 
\end{equation}

\noindent
Using the inequality 
$|\bar x|^2 \le \overline{|x|^2}$,
for a random variable $x_k$,
where equality is obtained only if
all the $x_k$'s are equal,
we find that
$P_{\rm max}^A \ge
P_{\rm max}^{AB}$.
This means that for a register in an entangled state
$| \psi \rangle$,
reducing the search space increases the success probability.
Equality is obtained only
if there is no entanglement between the systems held by Alice
and Bob, and Bob's system is in the equal superposition state
$| \eta \rangle_B$.
Thus, one can always add or remove unentangled qubits in the
state
$(|0\rangle +|1\rangle)/\sqrt{2}$
without changing 
$P_{\rm max}$.
However, adding entangled qubits reduces
$P_{\rm max}$
while removing such qubits increases it.

\subsection{Search using an initial pseudo-pure state}

Consider a search using an initial state which is a pseudo-pure
state of $n$ qubits

\begin{equation}
\rho_{\epsilon} = (1-\epsilon) \frac{I}{N}  
                + \epsilon |\psi\rangle \langle \psi|,  
\end{equation}

\noindent
where $\hat I$ is the $N$-dimensional unit matrix representing
the maximally mixed state, $|\psi\rangle$ is a pure state
and 
$0<\epsilon<1$.
The maximally mixed state can be represented by

\begin{equation}
I = \frac{1}{N} \sum_{i=0}^{N-1} |i\rangle \langle i|,
\end{equation}

\noindent
while the pure state
$| \psi \rangle = a_i |i\rangle$,
has an average amplitude
$\bar{a}_{\psi}$.
Using Eq. 
(\ref{<pst>_mix})
we find that

\begin{equation}
P_{\rm max} = \frac{1-\epsilon}{N} 
                       + \epsilon N |\bar a_{\psi}|^2
                       +O(\epsilon {r/N}).  
\end{equation}

\noindent
Therefore, 
the success probability is simply reduced by a 
factor of $\epsilon$ vs. its value in the case
that the initial state is the pure
state
$|\psi\rangle$.

\section{Summary}

We have introduced an algebraic approach to 
the analysis of 
Grover's quantum search algorithm
with an arbitrary initial quantum state.
This approach reveals the geometrical structure of the search space,
which turns out to be a four dimensional subspace of the
Hilbert space.
This approach 
unifies and generalizes earlier results on the 
time evolution of the amplitudes during the search,
the optimal number of iterations
and the success probability.
Furthermore, it enables a direct generalization to the case
in which the initial state is a mixed state, providing
an exact formula for the success probability.

\appendix
\section{The average success probability}

The probability of success of the quantum search algorithm
after $t$ iterations,
for a given choice of the set of $r$ marked states
is given by
Eq. (\ref{Pst}).
Since the set of marked states is not known
(although their number, $r$, is specified) 
the actual success 
probability is the average 
$\langle P_s(t) \rangle$ 
over all possible choices of the set of marked
states:

\begin{eqnarray}
\langle P_s(t) \rangle & = & 
\langle P_0 \rangle + 
\frac{1}{2} \left[(N-r)\langle {|{\bar{a}}_U|}^2 \rangle  - 
r \langle {|{\bar{a}}_M|}^2 \rangle \right]
\nonumber \\ 
& & - \frac{1}{2} \left[(N-r) \langle {|{\bar{a}}_U|}^2 \rangle 
-r \langle {|{\bar{a}}_M|}^2 \rangle \right] \cos(2\omega t)
\nonumber \\
& & + \frac{1}{2} \sqrt{r(N-r)} 
\left(\langle {{\bar{a}}_U}^*{\bar{a}}_M \rangle + 
\langle {{\bar{a}}_M}^*{\bar{a}}_U \rangle \right) \sin(2\omega t). 
\label{Pst_ave}
\end{eqnarray}

\noindent
We will now 
evaluate the averages 
$\langle P_0 \rangle$, 
$\langle {|{\bar{a}}_U}|^2 \rangle$, 
$\langle {|{\bar{a}}_M}|^2 \rangle$ 
and 
$\langle {{\bar{a}}_U}^*{\bar{a}}_M \rangle$. 
The average $\bar a$
over all
amplitudes 
in 
$|\psi \rangle$,
satisfies
the inequality
${|\bar{a}|}^2  \le {1}/{N}$ 
where equality is obtained only for the equal superposition
state
$|\eta\rangle$.
The average of
$P_0$
over all possible choices of the $r$ marked states
is

\begin{equation}
\langle P_0 \rangle = \langle \sum_{m \in {\cal M}} {|a_m|}^2 \rangle 
= \frac{C_{r-1}^{N-1}}{C_{r}^{N}} \sum_{j=0}^{N-1} {|a_j|}^2 = \frac{r}{N}, 
\label{<P0>}
\end{equation}

\noindent
where the binomial coefficient

\begin{equation}
C_{r}^{N} = \frac{N!}{(N-r)!r!}
\label{C_alpha_beta}
\end{equation}

\noindent
is the number of different sets of $r$ objects 
that can be picked out of
$N$ distinguishable objects.
The second moments of the amplitude distribution are

\begin{eqnarray}
\langle {|\bar{a}_U|}^2 \rangle  &=&  
\frac{1}{{(N-r)}^2} 
\langle 
\sum_{u_1 \in {\cal U}}a_{u_1}\sum_{u_2 \in {\cal U}} {a_{u_2}}^* 
\rangle
= 
{|\bar{a}|}^2 + O\left( 1/N^2 \right) 
\nonumber \\
\langle {|\bar{a}_M|}^2 \rangle  &=&  
{|\bar{a}|}^2\left(1-\frac{1}{r}\right)+\frac{1}{rN}+ 
O\left( 1/N^2 \right) 
\nonumber \\
\langle  {{\bar{a}}_U}^*{\bar{a}}_M \rangle  &=& 
\frac{1}{r(N-r)} \langle \sum_{m \in {\cal M}}a_m
\sum_{u \in {\cal U}} {a_u}^*  \rangle  
=
{|\bar{a}|}^2 + O\left( 1/N^2 \right) 
\nonumber \\
\langle  {\bar{a}}_M{{\bar{a}}_U}^* \rangle
&=& {\langle  {{\bar{a}}_U}^*{\bar{a}}_M \rangle}^* = 
{|\bar{a}|}^2 + O\left( 1/N^2 \right). 
\label{<aM_aU>}
\end{eqnarray}

\noindent
Substitution of the above averages in Eq. 
(\ref{Pst_ave}) yields 

\begin{equation}
\langle P_s(t) \rangle  =   
N{|\bar{a}|}^2 {\sin}^2 \left[\omega(t+1/2)\right] 
+\frac{r}{N} \left(1 - N{|\bar{a}|}^2\right) 
+ O\left( 1/N \right), 
\label{<pst>_appendix}
\end{equation}

\noindent
In the limit of a large search space and 
$r \ll N$, the expression for the success probability
is reduced to

\begin{equation}
\langle P_s(t) \rangle =   N{|\bar{a}|}^2 {\sin}^2 [\omega (t+1/2)] 
+ O\left({r/N}\right).
\label{<pst>_r<<N_appendix}
\end{equation}

%\bibliography{qc}
%\bibliographystyle{prsty}

\newpage 
\clearpage

\end{document}